\documentclass[pra,twocolumn,superscriptaddress,notitlepage]{revtex4-1}
\usepackage{makeidx}
\usepackage[utf8]{inputenc}
\usepackage{graphicx}
\usepackage{amsmath}
\usepackage{mathrsfs}
\usepackage{graphicx}
\usepackage{subfigure}
\usepackage{dcolumn}
\usepackage{bm}
\usepackage{bbm}
\usepackage{color}
\usepackage{esint}
\usepackage{lineno}
\usepackage[breaklinks,
            colorlinks,
            urlcolor=blue,
            linkcolor=blue,
            anchorcolor=blue,
            citecolor=blue]{hyperref}

\begin{document}

\title{Suppressing Decoherence in Quantum Plasmonic Systems \\ by Spectral Hole Burning Effect}

\author{Jia-Bin You}
\email{you\_jiabin@ihpc.a-star.edu.sg}
\affiliation{Institute of High Performance Computing, A*STAR (Agency for Science, Technology and Research), 1 Fusionopolis Way, \#16-16 Connexis, Singapore 138632}

\author{Xiao Xiong}
\affiliation{Institute of High Performance Computing, A*STAR (Agency for Science, Technology and Research), 1 Fusionopolis Way, \#16-16 Connexis, Singapore 138632}

\author{Ping Bai}
\affiliation{Institute of High Performance Computing, A*STAR (Agency for Science, Technology and Research), 1 Fusionopolis Way, \#16-16 Connexis, Singapore 138632}

\author{Zhang-Kai Zhou}
\affiliation{State Key Laboratory of Optoelectronic Materials and Technologies, School of Physics, Sun Yat-sen University, Guangzhou 510275, China}

\author{Wan-Li Yang}
\affiliation{State Key Laboratory of Magnetic Resonance and Atomic and Molecular Physics, Wuhan Institute of Physics and Mathematics, Chinese Academy of Sciences, Wuhan 430071, China}

\author{Ching Eng Png}
\affiliation{Institute of High Performance Computing, A*STAR (Agency for Science, Technology and Research), 1 Fusionopolis Way, \#16-16 Connexis, Singapore 138632}

\author{Leong Chuan Kwek}
\affiliation{Centre for Quantum Technologies, National University of Singapore, 3 Science Drive 2, Singapore 117543}
\affiliation{MajuLab, CNRS-UNS-NUS-NTU International Joint Research Unit, UMI 3654, Singapore}
\affiliation{National Institute of Education and Institute of Advanced Studies,
Nanyang Technological University, 1 Nanyang Walk, Singapore 637616}
\affiliation{School of Electrical and Electronic Engineering Block S2.1, 50 Nanyang Avenue, Singapore 639798}

\author{Lin Wu}
\email{wul@ihpc.a-star.edu.sg}
\affiliation{Institute of High Performance Computing, A*STAR (Agency for Science, Technology and Research), 1 Fusionopolis Way, \#16-16 Connexis, Singapore 138632}

\begin{abstract}

Quantum plasmonic systems suffer from significant decoherence due to the intrinsically large dissipative and radiative dampings.
Based on our quantum simulations $via$ a quantum tensor network algorithm, we numerically demonstrate the mitigation of this restrictive drawback by hybridizing a plasmonic nanocavity with an emitter ensemble with inhomogeneously-broadened transition frequencies.
By burning two narrow spectral holes in the spectral density of the emitter ensemble, the coherent time of Rabi oscillation for the hybrid system is increased tenfold.
With the suppressed decoherence, we move one step further in bringing plasmonic systems into practical quantum applications.

\end{abstract}

\maketitle

\section{Introduction}

Plasmonic cavity quantum electrodynamics (QED) at nanoscale opens up an unprecedented avenue to extreme light-matter interactions at room temperature and in ambient conditions \cite{Santhosh2016,Chikkaraddy2016,PhysRevLett.118.237401}, where plasmonic nanocavities offer subwavelength, sub-diffraction and significant local field confinement \cite{NatPhys9.329,Xu:18,jacob_2012,Bozhevolnyi2017,NatPhot4.83,Fitzgerald2016,ZHOU20191}.
Recently, plasmonic systems (\textit{e.g.}, waveguides, metasurfaces) have emerged as a natural choice to build compact photonic integrated circuits operating at the nanoscale for various quantum applications, such as quantum information processing \cite{Stav1101,Altewischer2002,PhysRevLett.94.110501,PhysRevLett.106.020501,doi:10.1021/acsphotonics.7b00717,doi:10.1021/acsphotonics.7b01241,doi:10.1021/acs.nanolett.7b03372,XX2019,Lee_2013,Fakonas_2015} and quantum computing \cite{Bekenstein2020,doi:10.1002/adma.201703986,Calafell2019}.
Compared to current noisy intermediate-scale quantum (NISQ) chips operating in cryogenic temperature and at the microscale (\textit{e.g.}, superconducting qubits, trapped ions  \cite{Arute2019,Majer2007,doi:10.1063/1.5088164,Murali2020}), these nanophotonic circuits potentially enable an ultimate miniaturization of photonic components for quantum optics, and also mark an important step towards the long-term goal of room-temperature quantum computing \cite{Bekenstein2020,Schneider2020,doi:10.1002/adma.201703986}.

However, there is a major hurdle for realizing plasmonic quantum information processing and quantum computing -- the intrinsically large absorption in the metals results in the fairly large decay rate of plasmonic polaritons. These energy dissipation processes unavoidably induce decoherence in the system and limit its performance \cite{Lee_2013}.
One way to overcome the loss or the decoherence problem is to explore quantum plasmonic systems that are strongly coupled to an ensemble of inhomogeneously-broadened quantum emitters ($i.e.$, different transition frequency for each emitter).
Modifying the spectral density of the emitter ensemble by a frequency-selective bleaching technique leads to an increased transmission at the burned spectral hole of the selected frequency.
Such spectral hole burning (SHB) effect, based on collective dark states \cite{PhysRevLett.115.033601,NatPhot11.36,doi:10.1002/lpor.201600189}, was suggested in microwave cavity QED to reduce the dissipation of the polaritons and suppress the decoherence of a hybrid system beyond the limit set individually by the cavity or the emitter ensemble.

In this work, we investigate the SHB effect in a hybrid plasmonic system to mitigate the large intrinsic dampings of the plasmonic system, which is coupled to an emitter ensemble with transition frequencies distributed in a frequency comb \cite{PhysRevLett.121.133601}.
Different from the microwave cavity \cite{PhysRevLett.115.033601,NatPhot11.36,doi:10.1002/lpor.201600189}, the plasmonic cavity operates generally at the nanoscale and there is insufficient space for a large number of emitters to efficiently couple to the plasmonic nanocavities (see Appendix \ref{capacity}), thus the continuum model for the emitter ensemble in the thermodynamic limit ($N\rightarrow\infty$) invalidates.
To treat each emitter discretely and solve this many-body problem accurately beyond the linear and mean-field approximations, we perform quantum simulations by employing the quantum tensor network algorithms -- the matrix product state (MPS) algorithm \cite{SCHOLLWOCK201196,Verstraete2008,PhysRevLett.114.220601,PhysRevA.92.022116,PhysRevLett.121.227401,PhysRevB.98.165416} to calculate the transmission spectra, and the time-dependent variational principle (TDVP) \cite{PhysRevLett.107.070601,PhysRevB.94.165116} to solve the dynamics of the system in time domain. The Rabi oscillation of the hybrid plasmonic system is observed, and the coherent time is obtained to be 10 times higher than the original plasmonic system.
To substantiate our experimental proposal, we combine our quantum simulations with an electromagnetic field simulation to study the commonly used plasmonic nanocavity-gold nanoparticle (AuNP) dimer, taking into account the temperature effect \cite{yeshchenko2013temperature} and laser ablation effect \cite{doi:10.1002/adma.201103807,tarasenko2005laser,app9030363} on the AuNPs during hole burning. Some other practical issues that might impact the SHB effect are also studied numerically, such as different pumping schemes, non-ideal frequency comb configurations, randomly-distributed transition frequencies for emitters, and the size of the emitter ensemble.

\begin{figure}[tbp]
\includegraphics[width=8cm]{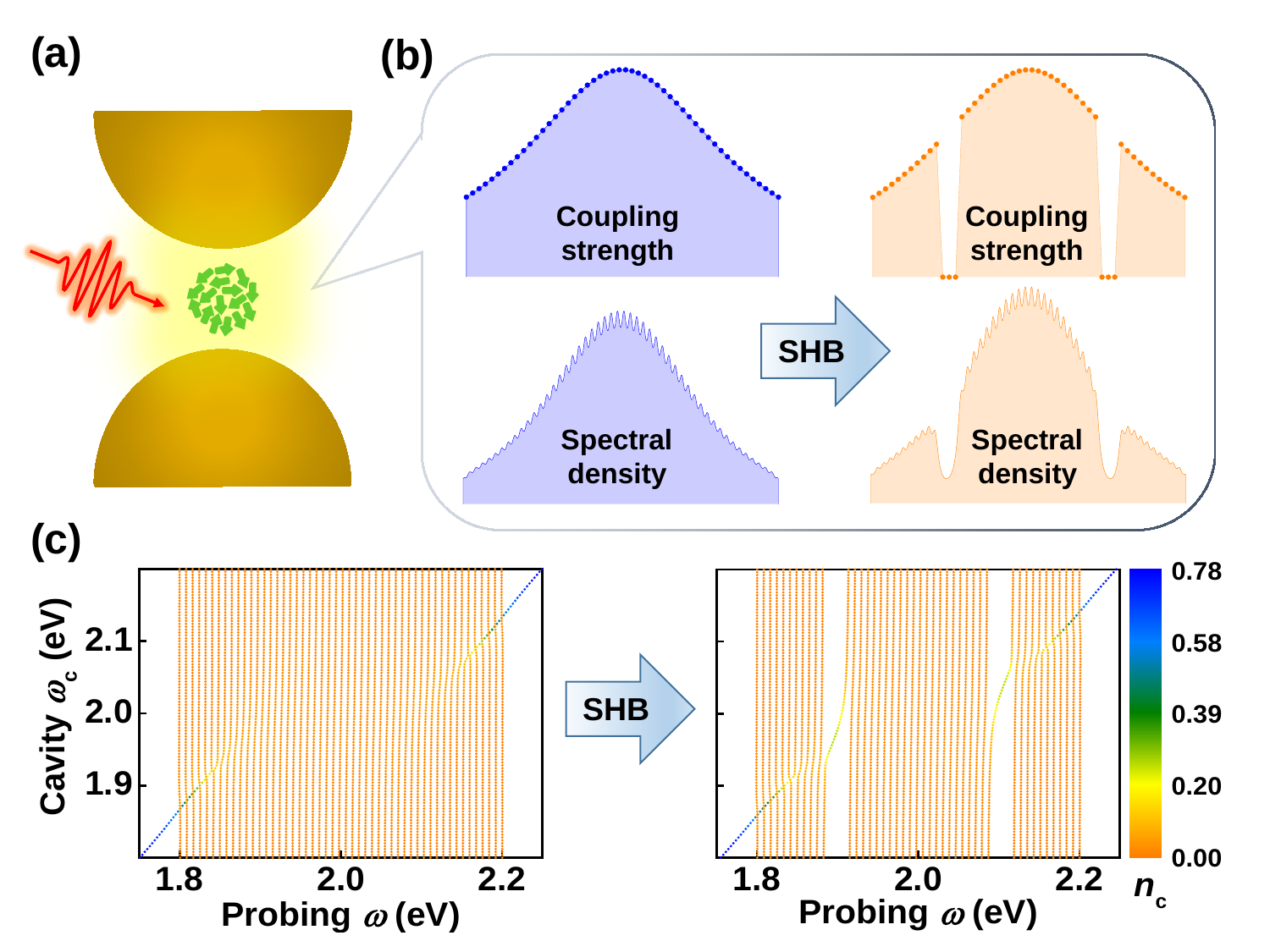}
\caption{
Spectral hole burning in hybrid plasmonic system.
(a) Schematic drawing of the hybrid system consisting of a plasmonic cavity coupled to an inhomogeneously-broadened emitter ensemble.
(b) Individual coupling strength between the cavity and each emitter in the ensemble, and the spectral density of the emitter ensemble before and after SHB.
(c) Single excitation spectra of the system before and after SHB, where the population of cavity photon $n_\textrm{c}$ is indicated by the colour bar.}
\label{fig1}
\end{figure}

\section{Spectral hole burning effect}

As illustrated in Fig. \ref{fig1}(a), a plasmonic nanocavity is coupled to an emitter ensemble that contains $N$ quantum emitters, each modeled as a two-level system with transition frequency $\omega_{i}$ and transition operator $\sigma_{i}^{+}=|\textrm{e}_{i}\rangle\langle{\textrm{g}_{i}}|$ between the ground state $|\textrm{g}_{i}\rangle$ and the excited state $|\textrm{e}_{i}\rangle$.
The plasmonic nanocavity, with resonant frequency $\omega_{\textrm{c}}$, is second quantized and described as a harmonic oscillator with the bosonic creation (annihilation) operator $a^{\dag} (a)$ with canonical commutation relation $[a,a^{\dag}]=1$. Each emitter couples to the plasmonic cavity mode through the Jaynes-Cummings interaction with a coupling strength $g_{i}$.
A driving laser field $E_{\textrm{L}}(t)$ with probing frequency $\omega$ pumps the entire system via the dipole moments of cavity $\mu_{\textrm{c}}$ and emitters $\mu_{\textrm{e}}$ with the strengths $\Omega_{\textrm{c}}(t)=\mu_{\textrm{c}}E_{\textrm{L}}(t)$ and $\Omega_{\textrm{e}}(t)=\mu_{\textrm{e}}E_{\textrm{L}}(t)$, respectively.
In a rotating frame with probing frequency $\omega$, the Hamiltonian of the system can be recast into $H=\Delta_{\textrm{c}}a^{\dag}a+\sum_{i=1}^{N}[\Delta_{i}\sigma_{i}^{+}\sigma_{i}^{-}+g_{i}(\sigma_{i}^{+}a+a^{\dag}\sigma_{i}^{-})]+\Omega_{\textrm{c}}(t)(a+a^{\dag})+\Omega_{\textrm{e}}(t)(S^{-}+S^{+})$,
where $S^{- (+)}=\sum_{i=1}^{N}\sigma_{i}^{-(+)}$, and $\Delta_{\textrm{c(i)}}=\omega_{\textrm{c(i)}}-\omega$ represents the detuning between the driving laser and the cavity (emitter). Since the energy of the cavity and emitters inevitably dissipates into the surrounding environment, the dynamics of such an open system is governed by the master equation \cite{PhysRevA.98.013839}, $\partial_{t}\rho=i[\rho,H]+\frac{\kappa}{2}\mathcal{D}[a]\rho+\frac{\Gamma_{i}}{2}\sum_{i}\mathcal{D}[\sigma_{i}^{-}]\rho$, where $\rho$ is the density matrix of the system and $\mathcal{D}[\hat{o}]\rho=2\hat{o}\rho\hat{o}^{\dag}-\hat{o}^{\dag}\hat{o}\rho-\rho\hat{o}^{\dag}\hat{o}$ is the Lindblad term that accounts for the losses from either cavity or emitter with decay rate $\kappa$ and $\Gamma_{i}$, respectively.

To demonstrate the SHB effect, we  first consider an ideal case where the emitters in the ensemble are arranged in a finite frequency comb \cite{PhysRevLett.121.133601} with transition frequencies spaced at equidistant intervals centering around $\omega_{\text{e}}$: $\omega_{i}=\omega_{\text{e}}-\Delta\omega+\tfrac{2\Delta\omega}{N-1}(i-1)$ in the range of $[\omega_{\text{e}}-\Delta\omega,\omega_{\text{e}}+\Delta\omega]$, with $i=1,2,...,N$.
The coupling strength between each emitter and the cavity follows Lorentzian distribution $g_{i}=\frac{A}{1+\beta(\omega_{i}-\omega_{\textrm{e}})^2}$ \cite{Umarov2008,PhysRevLett.115.033601}. This leads to the spectral density $\rho(\omega)$ of the emitter ensemble following $\Omega^2\rho(\omega)=\sum_{i}\frac{g_{i}^2}{(\Gamma_{i}/2)^2+\Delta_{i}^2}\Gamma_{i}$, as shown in the left panel of Fig. \ref{fig1}(b), where $\Omega^2=\sum_{i}g_{i}^2$ represents an effective coupling strength.
Throughout this study, the parameters $\omega_{\text{e}}=2$ eV, $\Delta\omega=0.2$ eV, $N=50$; $g_{i}=0\sim0.02$ eV (with $A=0.2$ eV, $\beta=0.1$);
$\Gamma_{i}=0.01$ eV, $\kappa=0.1$ eV; $\mu_{\textrm{c}}=19\mu_{\textrm{e}}$, and constant driving $\Omega_{\text{e}}=\Omega_{\text{c}}/19=0.001$ eV are used, except where otherwise stated.

Applying an intensive hole-burning pulse with intensity above a certain threshold on the emitter ensemble,  some emitters of selected frequency will be thermalized into an equal mixture of their ground and excited states, cancelling out their coherent light-matter interaction \cite{NatPhot11.36}, resulting in a zero coupling strength.
As exemplified in the right panel of Fig. \ref{fig1}(b), two spectral holes are symmetrically burned at $\omega_{\text{e}}\pm\Omega$ (or $\omega_{\text{e}}\pm0.102$ eV) in the spectral density $\rho(\omega)$ of ensemble,
which is equivalent to remove the emitters with position index $i=$ 12, 13, 14 and 37, 38, 39 in the comb,
thus resulting in two dips in the coupling strength spectrum with a width of $0.033$ eV.
We then restrict the total excitation to single-excitation subspace and calculate the population of the cavity photons, $n_{\textrm{c}}=\langle a^{\dag}a \rangle$, as a function of the probing frequency $\omega$.
By sweeping the resonant energy of cavity $\omega_{\textrm{c}}$, we plot these single-excitation energy spectra in Fig. \ref{fig1}(c).
Clearly, after hole burning, two states emerge within the spectral gaps and are isolated from the remaining subradiant states.
These two states are incorporated with the common ground states of the system, forming an effective V-level structure that naturally hosts the dark states \cite{RevModPhys.77.633}, which could potentially enhance the coherent time of the system.

\begin{figure}[tbp]
\includegraphics[width=8.5cm]{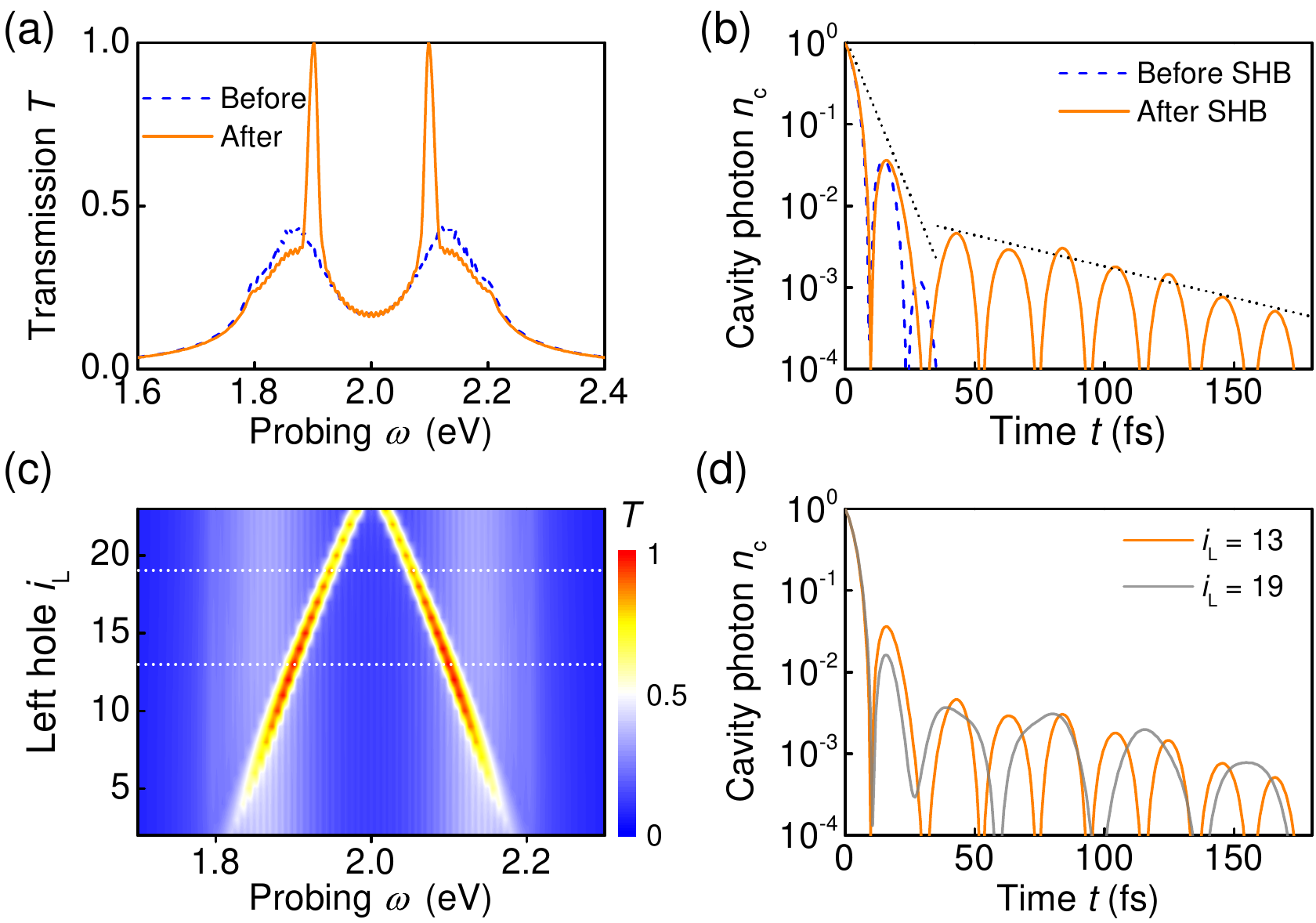}
\caption{The SHB effect on steady-state and dynamics of the hybrid plasmonic system.
(a) Normalized transmission spectrum and (b) Rabi oscillation of the cavity photon before and after SHB.
(c) Normalized transmission spectrum after SHB, as a function of the position of the left hole $i_{\textrm{L}}$, where the right hole is changed symmetrically.
(d) Tunable Rabi oscillations of the cavity photon for two different pairs of spectral holes:
$(i_{\textrm{L}},i_{\textrm{R}})=(13,38)$ and $(i_{\textrm{L}},i_{\textrm{R}})=(19,32)$.
In this study, $\omega_{c}=2$ eV, and the cavity (or all the emitters) is initially prepared in Fock state $|1\rangle$ (or in their ground states).}
\label{fig2}
\end{figure}

To fully understand the behavior of the long-lived dark states in this open quantum many-body system, we apply the variational MPS 
and the TDVP algorithms (see Appendix \ref{vMPS}) 
to study the steady state and the dynamics of the system.
Notice that the MPS approach is different from the mean-field solutions \cite{PhysRevLett.115.033601,NatPhot11.36,doi:10.1002/lpor.201600189}, which restricts the Hamiltonian to the single-excitation subspace (see Appendix \ref{comparison}).
We discuss the SHB effect by first plotting the normalized transmission spectrum of the plasmonic cavity $T(\omega)$ in Fig. \ref{fig2}(a), which is proportional to the scattered photon number $\langle{a^{\dag}a}\rangle$ from the cavity.
The SHB significantly modifies the emission of the hybrid plasmonic system.
Compared to the spectrum before SHB, two sharp peaks appear after the hole burning, which are the direct evidence of the collective dark states \cite{NatPhot11.36}.
This pair of dark states are well decoupled from the remaining subradiant states so that the broadening of the peaks (width of $0.011$ eV) remains relatively small compared to the background plasmonic polariton before SHB (decay rate $\kappa=0.1$ eV).
Here, the inhomogeneously-broadened lineshape $\rho(\omega)$ of the emitter ensemble corresponds to the superposition of many homogeneous lineshape of individual emitter $\frac{\Gamma_{i}}{(\Gamma_{i}/2)^2+\Delta_{i}^2}$ weighted by $g_{i}^2$ and shifted with each other in frequency space.
The burning laser will bleach the emitters that are nearly resonant with the laser ($g_{i}=0$) and create a spectral hole with lineshape corresponding to the bleaching emitters.
Therefore, the decay rate of the dark states is limited from below by the decay rate of a single emitter $\Gamma_{i}=0.01$ eV.
To further interpret the SHB effect, an analytical solution of the transmission is derived under linear and mean-field approximations in Appendix \ref{MFtheory}, where we get $T(\omega)\propto1/|\Delta_{\textrm{c}}-\Omega^2\delta(\omega)-i[\kappa+\Omega^2\rho(\omega)]/2|^2$ with Lamb shift $\delta(\omega)$. It is evident that the peaks of $T(\omega)$ appear at a probing frequency $\omega_{\textrm{peak}}$ where the denominator gets close to zero, that is, $\omega_{\textrm{c}}-\omega_{\textrm{peak}} = \Omega^2\delta(\omega_{\textrm{peak}})$ and $\rho(\omega_{\textrm{peak}})=0$.
This confirms that the transmission of the hybrid system can be tuned by modifying the properties of the emitter ensemble: Lamb shift $\delta(\omega)$ and spectral density $\rho(\omega)$.

The advantage of the hybrid system can also be remarked in the time-domain study as shown in Fig. \ref{fig2}(b), where we initially excite the cavity in the single-photon Fock state $|1\rangle$ and de-excite the emitters in their ground states.
After a background short-time Rabi oscillation, a long-lived oscillation resulting from the SHB gradually emerges.
The decay rate of the long-lived Rabi oscillation, which can be characterized by the slope, is about one magnitude smaller than that of the short-time Rabi oscillation. Therefore, the coherent time of the system is roughly 10-fold prolonged by the hole burning.

Interestingly, the pair of dark states is robust against a change of the hole burning positions as elaborated in Fig. \ref{fig2}(c), which can be observed widely from 1.8 to 2.2 eV spectrally. An optimal burning position centered at $(i_{\textrm{L}},i_{\textrm{R}})=(13,38)$ is found to achieve the most intensive peaks on top of the background spectrum, where $i_{\textrm{L}}$ and $i_{\textrm{R}}$ are the centers of the left and right hole positions.
Due to the small decay rate of the dark states, the Rabi splitting can even be observed in smaller spectral separation of hole positions centered at $(i_{\textrm{L}},i_{\textrm{R}})=(22,29)$.
In Fig. \ref{fig2}(d), we demonstrate the Rabi oscillations for two hole burning scenarios: $(i_{\textrm{L}},i_{\textrm{R}})=(13,38)$ and $(i_{\textrm{L}},i_{\textrm{R}})=(19,32)$, which are highlighted in the two white lines in Fig. \ref{fig2}(c).
It is observed that the period of Rabi oscillation is changed, implying that SHB not only suppresses the decoherence, but also allows us to control the Rabi frequency by varying the position of the spectral holes.

\begin{figure}[tbp]
\includegraphics[width=8cm]{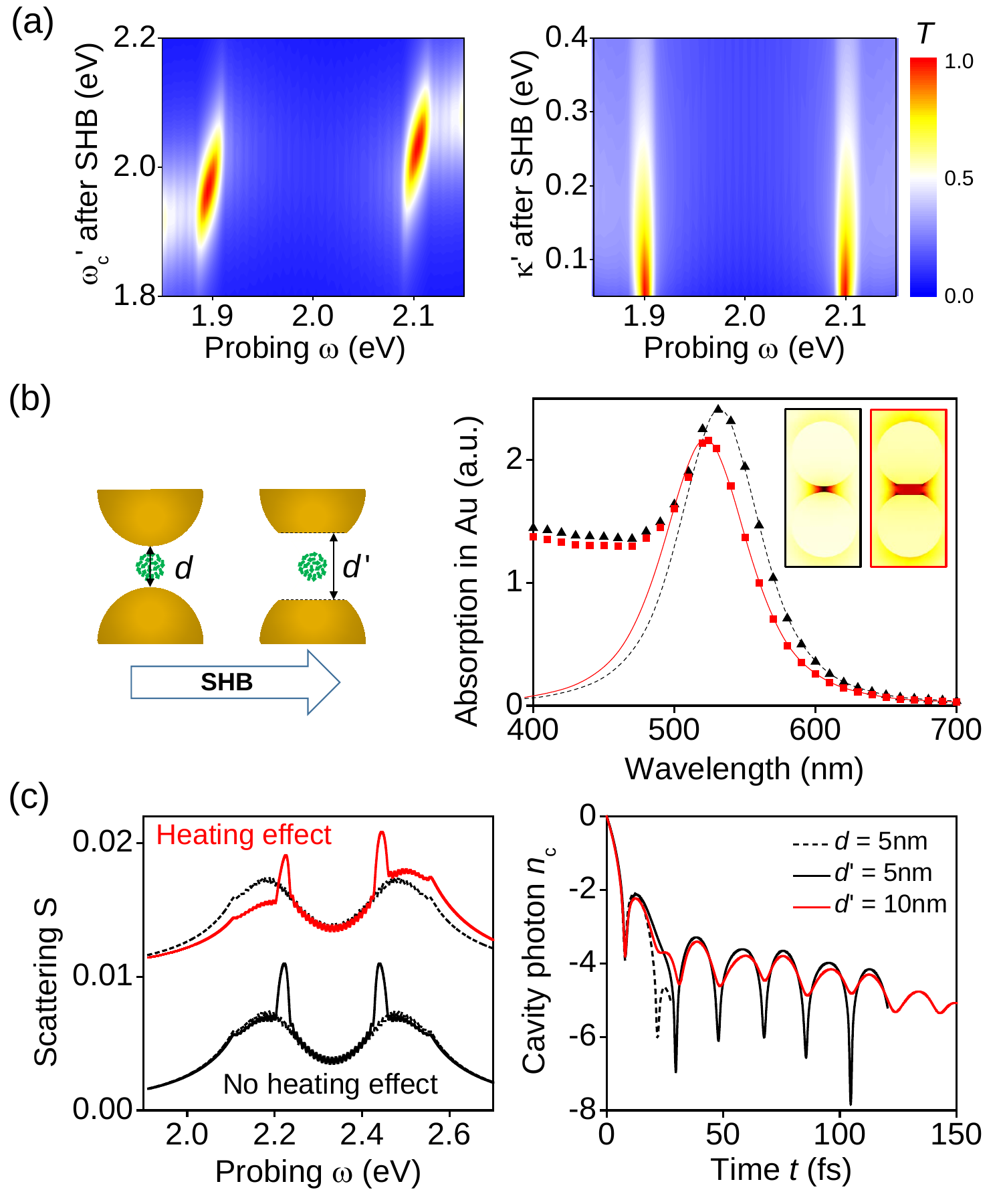}
\caption{
Heating effects on plasmonic nanocavity.
(a) General trend on varying $\omega_{\textrm{c}}'$ and $\kappa'$ after SHB.
(b)-(c) Case study on a AuNP dimer with diameter = 60 nm and $d=5$ nm: during SHB, the AuNP is assumed being partially ablated near the gap region hosting the emitter ensemble, modifying the shape of the AuNP and the gap spacing to $d'$.
(b) The full-wave simulations (symbols) to calculate the resonant frequency and the decay rate of the nanocavity, where the plasmon peak is Lorentz-fitted with (red solid) or without (black dashed) considering the heating effect on plasmonic nanocavity.
(c) Transmission spectrum and Rabi oscillation of the cavity photon before and after SHB, with or without considering the heating effects on plasmonic nanocavity.
}
\label{fig3}
\end{figure}

Despite its simplicity, our quantum many-body model provides good insights into more complex realistic experiments. We perform a few more simulations to study the role of randomness in the SHB effect. The detailed results are presented in Appendix \ref{randomness}, including
(i) non-ideal frequency comb,
(ii) a Lorentzian distributed ensemble with $N=5000$,
and (iii) different decay rates.
We find that the SHB effect is generally robust against the randomness in the emitter ensemble.

\section{Heating effects on plasmonic nanocavity}

Up to this point, we have focused on the emitter ensemble, and assumed that the plasmonic nanocavity is unaffected by the hole burning pulse that is intense enough to thermalize the emitter ensemble.
In reality, the burning pulse may induce local heating on the plasmonic metal nanoparticles, \textit{e.g.}, temperature effect \cite{yeshchenko2013temperature} or laser ablation effect \cite{doi:10.1002/adma.201103807,tarasenko2005laser,app9030363}, resulting in the changed properties of the plasmonic nanocavity ($i.e.$, $\omega_{\textrm{c}}$, and $\kappa$) during the hole burning process. We study the heating effects on the plasmonic nanocavity and their impacts on SHB.
As indicated in Fig. \ref{fig3}(a), when the plasmon resonance changes to $\omega_{\textrm{c}}'$  (either red-shift or blue-shift with respect to original $\omega_{\textrm{c}}=2$ eV), the two SHB peaks become asymmetric.
The clear feature of Rabi oscillation will gradually disappear when such shift exceeds 120 meV (see Appendix \ref{heating_time_domain}), defining the critical limit to observe SHB if plasmonic nanocavity is changed.
On the other hand,  the impact from the decay rate $\kappa'$ seems less critical. As expected, increased $\kappa'$ results in two blunt SHB peaks and reduced Rabi oscillation (see Appendix \ref{heating_time_domain}).

In Fig. \ref{fig3}(b)-(c), we demonstrate a realistic case study on a AuNP dimer (diameter = 60 nm and gap $d=5$ nm) with original $\omega_{\textrm{c}}=2.331$ eV and $\kappa=$ 0.287 eV. For such case, we select a resonant emitter ensemble ($\omega_{\textrm{e}}=2.331$ eV, $\Delta\omega=0.22$ eV) and burn two holes at $(i_{\textrm{L}},i_{\textrm{R}})=(13,38)$.
During SHB, the AuNP is assumed being partially ablated near the gap region hosting the emitter ensemble, modifying the shape of the AuNP and the gap spacing to $d'=10$ nm.
Despite such a huge change on the geometry of the nanocavity,
the resultant $\omega_{\textrm{c}}'=2.375$ eV and $\kappa'=0.313$ eV according to our full-wave simulation \cite{XX2019} in Fig. \ref{fig3}(b) only leads to a slightly varied SHB effect (upper red solid) in  Fig. \ref{fig3}(c). As compared to the SHB effect without the heating effect (lower black solid), the two SHB peaks become blunt. Meanwhile, this ablated plasmonic nanocavity only has small impact on the Rabi oscillation. This study further confirms the robustness of the SHB effect in practical scenarios.

\begin{figure}[tbp]
\includegraphics[width=8.5cm]{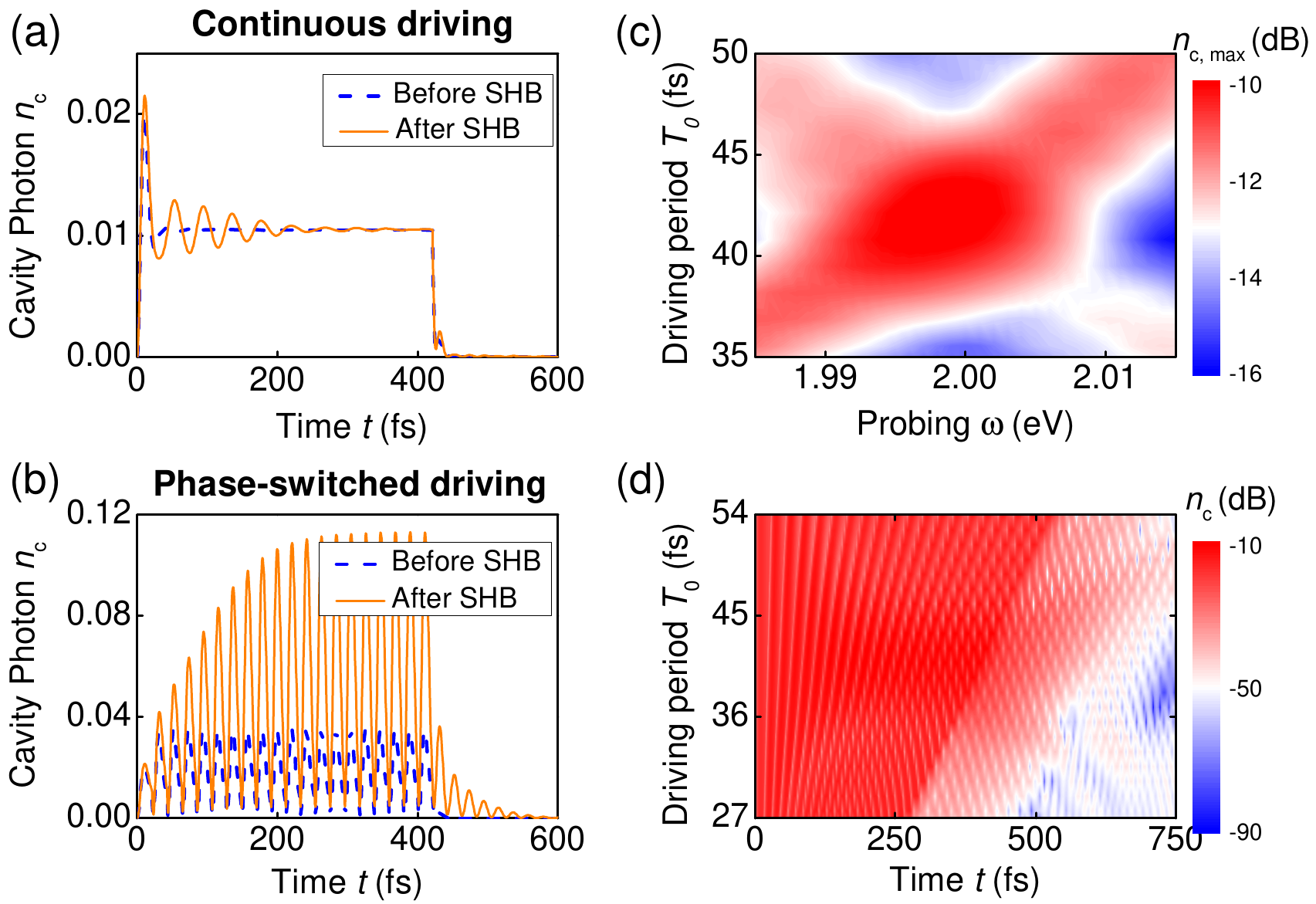}
\caption{
Different pumping schemes.
(a) Under a continuous driving, the dynamics of the cavity photons before and after SHB.
(b) Under a sequence of $\pi$-phase switched rectangular pulses (optimized setting: $T_0=42$ fs and $\omega=2$ eV), the dynamics of cavity photon with and without SHB.
(c) Contour plot in $\omega$--$T_0$ plane to obtain the most efficient rectangular pulses.
(d) Contour plot in the $t$--$T_0$ plane for fixed $\omega=2$ eV.
In this study, both the cavity and emitters are initially prepared in their ground states.
}
\label{fig4}
\end{figure}

\section{Different pumping schemes}

Finally, we consider different  pumping schemes and compare their potential performance.
Conventionally, it is usually to excite the system by a continuous wave that can be modeled as $E_{\textrm{L}}(t)=E_{\textrm{L}}$ (also the case in Figs. \ref{fig1} -- \ref{fig3}).
In Fig. \ref{fig4}(a), we study the quenching dynamics (how the system decays to its ground state) for the photon population in the cavity, after applying a continuous driving field until the system reaches the steady state, and then turning off the field.
Clearly, the cavity population decays rapidly once the driving is turned off either with or without SHB. In other words, the system is not excited efficiently even with SHB.
Alternatively, we can pump the system by a sequence of $\pi$-phase switched rectangular pulses \cite{PhysRevLett.115.033601}, $E_{\text{L}}(t)=\sum_{n=0}^{\infty}(-1)^{n}E_{\text{L}}[\Theta(t-nT_{0}/2)-\Theta(t-(n+1)T_{0}/2)]$, with a strength of $E_{\text{L}}$ and a period of $T_{0}$, and $\Theta(t)$ is the Heaviside function. As exemplified in Fig. \ref{fig4}(b), this procedure efficiently feeds energy into the hybrid system, leading to enhanced oscillation of cavity population. 
The maximal population for pulsed driving is a magnitude larger than that for continuous driving.
With SHB effect, not only the amplitude of driven oscillation is profoundly enhanced during the pumping, but also the relaxed oscillation after turning off the driving is dramatically longer-lived.
The amplitude of driven oscillation is found to be proportional to the square of driving strength (see Appendix \ref{dep_driving}).

In Fig. \ref{fig4}(c), we show how we can optimize this maximal population of the cavity against the driving period $T_{0}$ and the probing frequency $\omega$. An optimal population is found when the period is around the Rabi period, $2\pi/\Omega$ (= 42 fs) and the probing frequency is on resonance with the frequencies of the cavity and the emitters, $\omega=2$ eV. 
The SHB effect is also found robust against the driving period $T_0$ and the probing frequency $\omega$.
For instance, the SHB effect with various pulse periods is plotted in Fig. \ref{fig4}(d). Clearly, the SHB effect can be observed when the period ranges from 35 fs to 45 fs. 

\section{Conclusion}

We have theoretically demonstrated the SHB effect in hybrid plasmonic systems by quantum simulations using MPS and TDVP algorithms.
We show that the dissipation of the plasmonic polariton and the coherent time for the hybrid system can be significantly improved
by burning two narrow spectral holes in the spectral density of emitter ensemble with a frequency comb setup.
We also prove that the SHB effect can survive in randomness in potential experiments such as non-ideal frequency comb and Lorentzian-distributed emitter ensemble.
To substantiate the experimental justification of our proposal, we combine a full-wave electromagnetic field simulation into the quantum simulation to demonstrate SHB in a hybrid system consisting of a AuNP dimer coupled to an emitter ensemble, taking into account the heating effects of the hole burning pulse on AuNPs.
Finally, we suggest to drive the system using a sequence of $\pi$ phase-switched rectangular pulses, which can efficiently excite the system and further prolong the coherent Rabi oscillation.

\begin{acknowledgments}
The IHPC A*STAR Team acknowledges
the support from the National Research Foundation Singapore
(NRF2017-NRF-NSFC002-015 and QEP-SF1) and A*STAR Career Development Award (SC23/21-8007EP). W.-L. Yang acknowledges financial supports from the Youth Innovation Promotion Association (CAS No. 2016299).
\end{acknowledgments}

\appendix

\begin{widetext}

\section{Capacity of hosting emitters}
\label{capacity}

The capacity of hosting emitters depends on what kinds of plasmonic nanostructure are used. Here we give a brief survey on the number of emitters (J-aggregates) $N$ in the strong coupling hybrid plasmonic nanostructures.

\begin{table}[h]
\begin{tabular}{ |c|c|c| }
\hline
Plasmonic nanostructure & $N$ of J-aggregates & Reference \\
\hline\hline
Au nanovoids array & $1.6\times10^{6}$ & Ref. \cite{PhysRevLett.97.266808} \\
\hline
Au nanoslit array & $2000$ & Ref. \cite{Parinda2013} \\
\hline
Individual Au dimer & $203\sim614$ & Ref. \cite{doi:10.1021/nl4014887} \\
\hline
Ag triangular nanoprisim ensemble & 174 & Ref. \cite{Balci:13} \\
\hline
Single Ag nanorod & 110 & Ref. \cite{Gulis2013} \\
\hline
Single Ag triangular nanoprisim & $70\sim85$ & Ref. \cite{PhysRevLett.114.157401} \\
\hline
Single NPoM nanostructure& $1\sim10$ & Ref. \cite{Chikkaraddy2016} \\
\hline
Single cuboid Au@Ag nanorod & $1\sim7$ & Ref. \cite{PhysRevLett.118.237401} \\
\hline
\end{tabular}
\caption{The capacity of hosting emitters in different plasmonic nanostructures.}
\end{table}

Thus for plasmonic nanoarrays, we can use a large emitter ensemble ($N\sim10^3$) randomly sampled from the Lorentzian distribution. However, for the single plasmonic nanostructure such as AuNP dimer in Fig. 1(a) in the maintext, only small ensemble ($N\sim10^{1-2}$) can be applied as the hotspot of the gap mode hosts less emitters.

\section{Matrix product state algorithm for Tavis-Cummings model}
\label{vMPS}

\subsection{Ground state search}
\label{gs_search}

In this section, we will discuss in details about the implementation of matrix product state (MPS) algorithm for Tavis-Cummings model (TCM). In a rotating frame with probing frequency $\omega$, the plasmonic cavity coupled by an emitter ensemble with $N$ quantum emitters can be modeled as a TCM model given by
\begin{equation}
\label{TCM}
\begin{split}
H=\Delta_{\textrm{c}}a^{\dag}a+\sum_{i=1}^{N}[\Delta_{i}\sigma_{i}^{+}\sigma_{i}^{-}+g_{i}(\sigma_{i}^{+}a+a^{\dag}\sigma_{i}^{-})]+\Omega_{\textrm{c}}(a+a^{\dag})+\Omega_{\textrm{e}}(S^{-}+S^{+}),\\
\end{split}
\end{equation}
where $a^{\dag}$ is the creation operator for the plasmonic mode following canonical commutation relation $[a,a^{\dag}]=1$, $\sigma_{i}^{+}=|\text{e}_{i}\rangle\langle{\text{g}_{i}}|$ is the raising operator between ground state $|\text{g}_{i}\rangle$ and excited state $|\text{e}_{i}\rangle$ of emitter $i$ and $S^{+}=\sum_{i=1}^{N}\sigma_{i}^{+}$.
The emitter $i$ couples to the plasmonic mode through the Jaynes-Cummings interaction with coupling strength $g_{i}$. The driving strengths for the cavity and emitter are given by $\Omega_{\textrm{c}}$ and $\Omega_{\textrm{e}}$, respectively. The laser detunings for the cavity and emitters are given by $\Delta_{\textrm{c}}=\omega_{\textrm{c}}-\omega$ and $\Delta_{i}=\omega_{i}-\omega$ where $\omega_{\textrm{c}}$ and $\omega_{i}$ are the resonant frequency for cavity and transition frequencies for emitter $i$.

MPS is a well-known and successful example of the tensor network family. It is very well-suited to study gapped 1D or quasi-1D quantum many body systems \cite{SCHOLLWOCK201196,Verstraete2008}. The MPS consists of one-dimensional array of tensors. Each tensor represents one site in the many body system and the tensors are connected together by the bond indices each of which can take up to $D$ different values. Another index corresponds to the physical index of each site which can take $d$ different values. For example $d=2$ for a quantum bit.

To implement the MPS algorithm, the many-body quantum state and Hamiltonian should be first transformed to the MPS and matrix product operator (MPO). The MPS for a quantum state of Eq. (\ref{TCM}) can be written as
\begin{equation}
\label{MPS}
\begin{split}
|\psi\rangle=\sum_{a_{1},\cdots,a_{N}=1}^{D}\sum_{s_{1}=1}^{d_{\textrm{c}}}\sum_{s_{2},\cdots,s_{N+1}=1}^{d_{s}}M_{1,a_{1}}^{s_{1}}M_{a_{1},a_{2}}^{s_{2}}{\cdots}M_{a_{N},1}^{s_{N+1}}|s_{1}s_{2}{\cdots}s_{N+1}\rangle,\\
\end{split}
\end{equation}
where the dimensions of ``on-site" tensors $M_{1,a_{1}}^{s_{1}},M_{a_{1},a_{2}}^{s_{2}},\cdots,M_{a_{N},1}^{s_{N+1}}$ are $1\times{D}\times{d_{\textrm{c}}}, D\times{D}\times{d_{s}},\cdots,D\times{1}\times{d_{s}}$, respectively. Here $D$ is the maximum bond dimension and $d_{\textrm{c}},d_{s}$ are the physical dimension of cavity and atom, respectively. Based on the form of MPS, we note that the Hamiltonian of Eq. (\ref{TCM}) can be interpreted as a one-dimension model with long-range interaction between cavity and each emitter. This will lead to the particle number $N$ dependence of the bond dimension of MPO for the Hamiltonian. In the case of TCM in Eq. (\ref{TCM}), there is $N+1$ sites in the model and the bond dimension of the MPO for the Hamiltonian is $2(N+1)$. Take $N=2$ as an example. By expressing the bond indices explicitly, the ``on-site" MPO tensors for cavity and atom sites can be written as:
\begin{equation}
\begin{split}
W^{s_{1}s'_{1}}=&\left[\begin{array}{*{20}ccccccccc}
{\Delta_{\textrm{c}}a^{\dag}a+\Omega_{\textrm{c}}(a+a^{\dag})} & {g_{1}a} & {g_{1}a^{\dag}} & {g_{2}a} & {g_{2}a^{\dag}} & {I}\\
\end{array}\right],\\
\end{split}
\end{equation}
\begin{equation}
\begin{split}
W^{s_{i}s'_{i}}=&\left[\begin{array}{*{20}ccccccccc}
{I} & {0} & {0} & {0} & {0} & {0}\\
{\sigma_{i}^{+}} & {0} & {0} & {0} & {0} & {0}\\
{\sigma_{i}^{-}} & {0} & {0} & {0} & {0} & {0}\\
{0} & {I} & {0} & {0} & {0} & {0}\\
{0} & {0} & {I} & {0} & {0} & {0}\\
{\Delta_{i}\sigma_{i}^{+}\sigma_{i}^{-}+\Omega_{\textrm{e}}(\sigma_{i}^{-}+\sigma_{i}^{+})} & {0} & {0} & {0} & {0} & {I}\\
\end{array}\right], (i=2,3,\cdots,N),\\
\end{split}
\end{equation}
\begin{equation}
\begin{split}
W^{s_{N+1}s'_{N+1}}=&\left[\begin{array}{*{20}ccccccccc}
{I}\\
{\sigma_{N+1}^{+}}\\
{\sigma_{N+1}^{-}}\\
{0}\\
{0}\\
{\Delta_{N+1}\sigma_{N+1}^{+}\sigma_{N+1}^{-}+\Omega_{\textrm{e}}(\sigma_{N+1}^{-}+\sigma_{N+1}^{+})}\\
\end{array}\right].\\
\end{split}
\end{equation}
Therefore, the MPO form for the Hamiltonian of Eq. (\ref{TCM}) is
\begin{equation}
\begin{split}
H=\sum_{b_{1},\cdots,b_{N}=1}^{2(N+1)}\sum_{s_{1}=1}^{d_{\textrm{c}}}\sum_{s'_{1}=1}^{d_{\textrm{c}}}\sum_{s_{2},\cdots,s_{N+1}=1}^{d_{s}}\sum_{s'_{2},\cdots,s'_{N+1}=1}^{d_{s}}W_{1,b_{1}}^{s_{1}s'_{1}}W_{b_{1},b_{2}}^{s_{2}s'_{2}}{\cdots}W_{b_{N},1}^{s_{N+1}s'_{N+1}}|s_{1}s_{2}{\cdots}s_{N+1}{\rangle\langle}s'_{1}s'_{2}{\cdots}s'_{N+1}|.
\end{split}
\end{equation}

To find the ground state, we can minimize the energy $E=\langle{\psi[\mathbf{M}]}|H|{\psi[\mathbf{M}]}\rangle$ subjected to the normalization condition $\langle{\psi[\mathbf{M}]}|{\psi[\mathbf{M}]}\rangle=1$. Here the variational MPS ansatz $|{\psi}[\mathbf{M}]\rangle$ is employed, where $\mathbf{M}=\{M^{s_{1}},M^{s_{2}},{\cdots}M^{s_{N+1}}\}$. By the method of Lagrange multipliers, the local minimization for site $k$ is equivalent to the following equation:
\begin{equation}
\begin{split}
\partial_{M_{a_{l},a_{l-1}}^{s_{k}*}}[\langle{\psi[\mathbf{M}]}|H|{\psi[\mathbf{M}]}\rangle-\lambda(\langle{\psi[\mathbf{M}]}|{\psi[\mathbf{M}]}\rangle-1)]=0,\\
\end{split}
\end{equation}
which leads to
\begin{equation}
\label{optimization}
\begin{split}
&\sum_{\{a'_{i},b_{i},s'_{j}\}}\sum_{\{a_{i\ne{l-1,l}},s_{j\ne{k}}\}}(M_{a_{1},1}^{s_{1}*}W_{1,b_{1}}^{s_{1}s'_{1}}M_{1,a'_{1}}^{s'_{1}})\cdots(W_{b_{i-1},b_{i}}^{s_{k}s'_{j}}M_{a'_{i-1},a'_{i}}^{s'_{j}}){\cdots}(M_{1,a_{N}}^{s_{N+1}*}W_{b_{N},1}^{s_{N+1}s'_{N+1}}M_{a'_{N},1}^{s'_{N+1}})\\
&-\lambda\sum_{\{a'_{i}\}}\sum_{\{a_{i\ne{l-1,l}},s_{j\ne{k}}\}}(M_{a_{1},1}^{s_{1}*}M_{1,a'_{1}}^{s_{1}})\cdots(M_{a'_{i-1},a'_{i}}^{s_{k}}){\cdots}(M_{1,a_{N}}^{s_{N+1}*}M_{a'_{N},1}^{s_{N+1}})=0.\\
\end{split}
\end{equation}
Notice that if we express the MPS of Eq. (\ref{MPS}) in the mixed-canonical form \cite{SCHOLLWOCK201196}, Eq. (\ref{optimization}) can be further reduced to an eigenproblem,
\begin{equation}
\label{eigenproblem}
\begin{split}
\sum_{a'_{l-1},a'_{l},s'_{k}}H_{a_{l-1},a_{l},a'_{l-1},a'_{l}}^{s_{k},s'_{k}}M_{a'_{l-1},a'_{l}}^{s'_{k}}={\lambda}M_{a_{l-1},a_{l}}^{s_{k}},\\ \end{split}
\end{equation}
where the effective Hamiltonian at site $k$ is
\begin{equation}
\begin{split}
H_{a_{l-1},a_{l},a'_{l-1},a'_{l}}^{s_{k},s'_{k}}=\sum_{\{a_{i\ne{l-1,l}},s_{j\ne{k}},b_{i},a'_{i\ne{l-1,l}},s'_{j\ne{k}}\}}(M_{a_{1},1}^{s_{1}*}W_{1,b_{1}}^{s_{1}s'_{1}}M_{1,a'_{1}}^{s'_{1}})\cdots(W_{b_{i-1},b_{i}}^{s_{k}s'_{k}}){\cdots}(M_{1,a_{N}}^{s_{N+1}*}W_{b_{N},1}^{s_{N+1}s'_{N+1}}M_{a'_{N},1}^{s'_{N+1}}).\\
\end{split}
\end{equation}
By solving for the lowest eigenvalue $\lambda_{\text{min}}$ and the corresponding eigenvector $M_{a_{l-1},a_{l}}^{s_{k}}$ of Eq. (\ref{eigenproblem}), we obtain the current ground state energy estimate. Therefore, the ground state search can be iteratively obtained by the sweep algorithm. For TCM of Eq. (\ref{TCM}), we sweep forward from $1$ to $N+1$ and backward from $N+1$ to $1$ for several times until the local lowest energy $\lambda_{\text{min}}$ converges to the ground state energy $E_{\text{g}}$ and the corresponding ground state $\psi[\mathbf{M_{\text{g}}}]$ is obtained.

\subsection{Time evolution by time-dependent variational principle}
\label{TDVP}

The Dirac-Frenkel time-dependent variational principle (TDVP), has been reformulated for the variational MPS \cite{PhysRevLett.107.070601}. The key ingredient is to project the right-hand side of the time-dependent Schr\"{o}dinger equation, $H\psi[\mathbf{M}]$, onto the tangent space, so that the evolution never leaves the manifold. This approach is independent of the Hamiltonian and can be implemented efficiently for long-range Hamiltonian. Concretely, it approximates the time evolution of an MPS $\psi[\mathbf{M}]$ under the Hamiltonian $H$ by minimizing
\begin{equation}
\begin{split}
\min_{\dot{\mathbf{M}}}\left|i\dot{\mathbf{M}}\partial_{\mathbf{M}}\psi[\mathbf{M}]-H\psi[\mathbf{M}]\right|^2
\end{split}
\end{equation}
with $\psi[\mathbf{M}]$ kept fixed while its derivative $\dot{\mathbf{M}}$ is varied.

More recently, an improved TDVP algorithm was derived for finite MPS with open boundaries, which relies on the mixed canonical gauge \cite{PhysRevB.94.165116,PhysRevB.101.235123}. This approach leads to an effective Schr\"{o}dinger equation for states constrained to the MPS manifold,
\begin{equation}
\begin{split}
i\frac{d}{dt}\psi[\mathbf{M}(t)]=P_{T\psi}H\psi[\mathbf{M}(t)],
\end{split}
\end{equation}
where $P_{T\psi}$ is an orthogonal projector onto the tangent space of $\psi[\mathbf{M}(t)]$. For the TCM in Eq. (\ref{TCM}), the tangent space projector can be decomposed as
\begin{equation}
\begin{split}
P_{T\psi}=\sum_{i=1}^{N+1}P_{L}^{[1:i-1]}\otimes{I_{i}}{\otimes}P_{R}^{[i+1:N+1]}-\sum_{i=1}^{N}P_{L}^{[1:i]}{\otimes}P_{R}^{[i+1:N+1]},\\
\end{split}
\end{equation}
where
\begin{equation}
\begin{split}
P_{L}^{[1:i-1]}&=\sum_{k=1}^{D}|\Phi_{L,k}^{[1:i-1]}{\rangle\langle}\Phi_{L,k}^{[1:i-1]}|,\\
P_{R}^{[i+1:N+1]}&=\sum_{k=1}^{D}|\Phi_{R,k}^{[i+1:N+1]}{\rangle\langle}\Phi_{R,k}^{[i+1:N+1]}|,\\
\end{split}
\end{equation}
meaning that
\begin{equation}
\begin{split}
|\psi[\mathbf{M}(t+\delta{t})]\rangle=\exp(-i\delta{t}P_{T\psi}H)|\psi[\mathbf{M}(t)]\rangle
\end{split}
\end{equation}
can be approximated by applying a Lie-Trotter-Suzuki decomposition \cite{Hatano2005} to the exponential. Here $|\Phi_{L,k}^{[1:i]}\rangle$ and $|\Phi_{R,k}^{[i+1:N+1]}\rangle$ are obtained by bipartitioning the TCM model into sites $[1:i]$ and $[i+1:N+1]$ and performing the Schmidt decomposition
\begin{equation}
\begin{split}
|\psi[\mathbf{M}]\rangle=\sum_{k=1}^{D}\lambda_{k}|\Phi_{L,k}^{[1:i]}{\rangle}\otimes|\Phi_{R,k}^{[i+1:N+1]}\rangle.\\
\end{split}
\end{equation}
Consequently, one can sweep back and forth along the MPS, time evolving one site tensor at a
time. This algorithm is symplectic and conserves the energy and norm of a state.

\subsection{Variational MPS algorithms for Lindblad master equation}

In general the emitters and the plasmonic cavity are lossy, which arises from spontaneous emission, imperfections in the cavity, and non-radiative losses due to the larger environment. These need to be accounted for in the description of the system \cite{PhysRevLett.121.133601}. In a Markovian setting, such losses in an open system can be described by using a Lindblad master equation of the form
\begin{equation}
\begin{split}
\partial_{t}\rho&=i[\rho,H]+\frac{\kappa}{2}\mathcal{D}[a]\rho+\sum_{i=1}^{N}\frac{\Gamma_{i}}{2}\mathcal{D}[\sigma_{i}^{-}]\rho,\\
\end{split}
\end{equation}
where $\mathcal{D}[\hat{o}]\rho=2\hat{o}\rho\hat{o}^{\dag}-\{\hat{o}^{\dag}\hat{o},\rho\}$.
It is straightforward to find that the master equation can be rewritten into
\begin{equation}
\begin{split}
\partial_{t}\rho&=i({\rho}H_{\text{eff}}^{\dag}-H_{\text{eff}}\rho)+\kappa{a}\rho{a}^{\dag}+\sum_{i=1}^{N}\Gamma_{i}\sigma_{i}^{-}\rho\sigma_{i}^{+},\\
\end{split}
\end{equation}
where
\begin{equation}
\begin{split}
H_{\text{eff}}=(\Delta_{\textrm{c}}-i\kappa/2)a^{\dag}a+\sum_{i=1}^{N}[(\Delta_{i}-i\Gamma_{i}/2)\sigma_{i}^{+}\sigma_{i}^{-}+g_{i}(\sigma_{i}^{+}a+a^{\dag}\sigma_{i}^{-})]+\Omega_{\textrm{c}}(a+a^{\dag})+\Omega_{\textrm{e}}(S^{-}+S^{+}).\\
\end{split}
\end{equation}
In the Choi representation \cite{PhysRevLett.114.220601,PhysRevA.92.022116,PhysRevLett.121.133601}, the master equation can be recast into $\partial_{t}|{\rho}\rangle\rangle=\mathcal{L}|{\rho}\rangle\rangle$, which has great similarity with the time-dependent Schr\"{o}dinger equation shown in Subsection (\ref{TDVP}). Here the density matrix $\rho$ is reshaped into a column vector $|{\rho}\rangle\rangle$ by concatenating all its columns and the Liouvillian superoperator is reformulated to operate on the corresponding enlarged Hilbert space as
\begin{equation}
\begin{split}
\mathcal{L}&=i(H_{\text{eff}}^{*}\otimes{I}-I\otimes{H_{\text{eff}}})+\kappa{a}\otimes{a}+\sum_{i=1}^{N}\Gamma_{i}\sigma_{i}^{-}\otimes\sigma_{i}^{-},\\
\end{split}
\end{equation}
then the variational MPS algorithms described in Subsections (\ref{gs_search}) and (\ref{TDVP}) can be applied to study the steady and dynamical properties of the system. The determination of the steady density matrix can be reformulated as the variational minimization \cite{PhysRevA.92.022116} of the Euclidean norm functional $|\mathcal{L}|{\rho}\rangle\rangle|\ge0$ and the time evolution of the system can be achieved by the TDVP algorithm \cite{PhysRevB.94.165116}. Notice that the expectation value of an observable $\hat{O}$ is $\langle\hat{O}\rangle=\text{tr}(\hat{O}\rho)=\langle\langle\hat{O}^{\dag}|\rho\rangle\rangle$ in Choi's representation.

\section{Comparison between the mean-field and matrix product state calculations}
\label{comparison}

\begin{figure}[h]
\includegraphics[width=6cm]{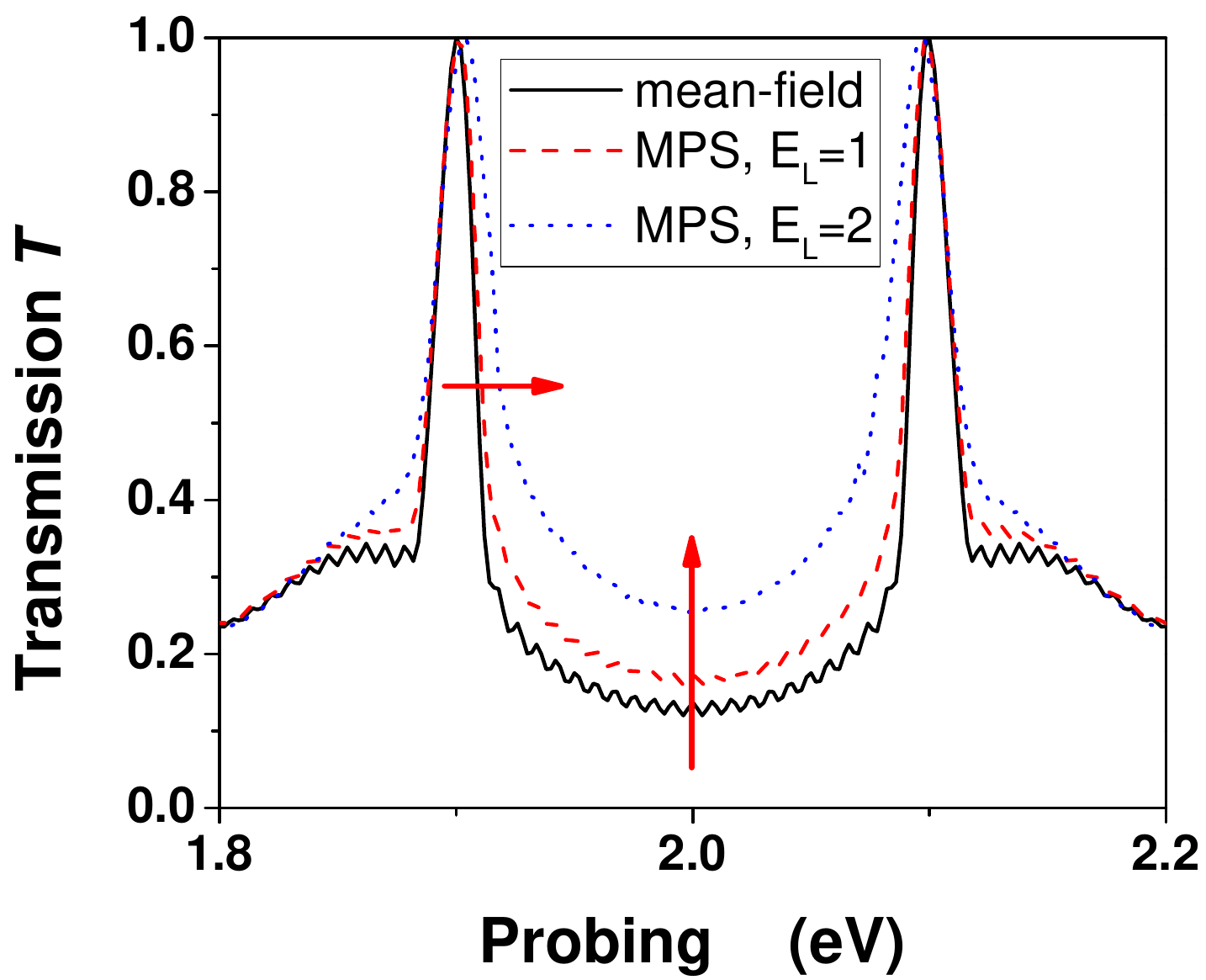}
\caption{Comparison between the mean-field and matrix product state calculations. The black solid line are the result from mean-field calculation. The red dashed and blue dotted lines are for driving strength $E_{\textrm{L}}=1$ and $E_{\textrm{L}}=2$, respectively. Other parameters used are kept the same as those in Fig. 2(a) in the main text.}
\label{figs7}
\end{figure}

Here we compare the mean-field with the matrix product state calculations. A driving laser field $E_{\textrm{L}}$ with probing frequency $\omega$ pumps the entire system via the dipole moments of cavity $\mu_{\textrm{c}}$ and emitters $\mu_{\textrm{e}}$ with the strengths $\Omega_{\textrm{c}}=\mu_{\textrm{c}}E_{\textrm{L}}$ and $\Omega_{\textrm{e}}=\mu_{\textrm{e}}E_{\textrm{L}}$. In Fig. \ref{figs7}, we find that the mean field solution is only exact when the driving strength $E_{\textrm{L}}$ is small, the peak-to-background ratio and the line profile of SHB peak will become lower and broader when one increases the driving strength $E_{\textrm{L}}$.

\section{Analytical solution of transmission}
\label{MFtheory}

An analytical solution of the transmission $T$ is derived to assist interpreting the spectral hole burning (SHB) effect.
We consider a simplified case where the coherent driving field only acts on the plasmonic cavity, $\Omega_{\textrm{e}}(t)=0$.
In the limit of low driving intensity when the linear approximation, $\langle{\sigma_{i,z}}\rangle\approx-1$, is valid, 
we can derive the equations of motion for the system as:
\begin{equation} \label{eom}
\begin{split}
\langle{\dot{a}}\rangle &= -(i\Delta_{\textrm{c}}+\kappa/2)\langle{a}\rangle-i\sum_{i}g_{i}\langle{\sigma_{i}^{-}}\rangle-i\Omega_{\textrm{c}}(t) \\
\langle{\dot{\sigma_{i}^{-}}}\rangle&=-(i\Delta_{i}+\Gamma_{i}/2)\langle{\sigma_{i}^{-}}\rangle-ig_{i}\langle{a}\rangle
\end{split}
\end{equation}
After some straightforward calculations, the transmission spectrum of the hybrid plasmonic system $T(\omega)$, proportional to the emitted photon number of the cavity, $\langle{a^{\dag}a}\rangle$, can be simplified to:
\begin{equation} \label{transmission}
T(\omega)\propto
\frac{1}{|\Delta_{\textrm{c}}-\Omega^2\delta(\omega)-i[\kappa+\Omega^2\rho(\omega)]/2|^2},
\end{equation}
where the mean-field approximation, $\langle{a^{\dag}a}\rangle\approx|\langle{a}\rangle|^2$, is applied. Here, $\Omega^2=\sum_{i}g_{i}^2$ represents an effective coupling strength that is enhanced by a factor of $\sqrt{N}$ compared to individual coupling strength $g_{i}$. The $\delta(\omega)$ and $\rho(\omega)$ represent the Lamb shift \cite{PhysRevLett.115.033601} and the spectral density of the emitter ensemble, respectively.

Looking at the denominator of this analytical solution of $T(\omega)$,  we can clearly see that the resonant frequency and the spectral broadening of the plasmonic cavity are modified by the dressed emitter ensemble.
In particular, the resonant frequency $\omega_{\textrm{c}}$ is shifted by the Lamb shift \cite{PhysRevLett.115.033601}, $\Omega^2\delta(\omega)=\sum_{i}\frac{g_{i}^2}{(\Gamma_{i}/2)^2+\Delta_{i}^2}\Delta_{i}$,
whereas the spectral broadening $\kappa$ is increased by the density of states of the emitter ensemble,  $\Omega^2\rho(\omega)=\sum_{i}\frac{g_{i}^2}{(\Gamma_{i}/2)^2+\Delta_{i}^2}\Gamma_{i}$. More importantly, the  value of $T(\omega)$ can be maximized when the denominator gets close to zero, that is, $\Delta_{\textrm{c}} = \Omega^2\delta(\omega)$ and $[\kappa+\Omega^2\rho(\omega)]/2=0$.
This implies that we could tune the transmission spectrum of the hybrid system by modifying the properties of the emitter ensemble $\delta(\omega)$ and $\rho(\omega)$. This analytical solution has been used in interpreting Fig. 2 in the main text.

\section{Randomness in spectral hole burning effect}
\label{randomness}



\subsection{Nonideal frequency comb}
Nonideal frequency comb refers to the case that the emitter frequencies are not exactly located at the comb position.
This can be modeled as a disorder among the transition frequencies of the ideal comb, where the on-site energy of the emitters becomes $H_{\textrm{e}}=\sum_{i}\omega'_{i}\sigma_{i}^{+}\sigma_{i}^{-}$.
The nonideality is reflected in $\omega'_{i}=\omega_{i}+\delta\omega_{i}$,
where $\delta\omega_{i}=\alpha_{i}\Delta\tfrac{\omega}{N-1}$ is a random on-site energy and the random number $\alpha_{i}\in[-r,r]\ (0\le{r}<1)$ is uniformly distributed. Meanwhile, the corresponding coupling strength for each modified transition frequency follows the same Lorentzian distribution $g'_{i}=\frac{A}{1+\beta(\omega'_{i}-\omega_{\textrm{e}})^2}$..
As shown in Fig. \ref{figs1}, we find that the SHB effect can still be observed in the presence of this nonideal frequency comb.

\begin{figure}[h]
\includegraphics[width=6cm]{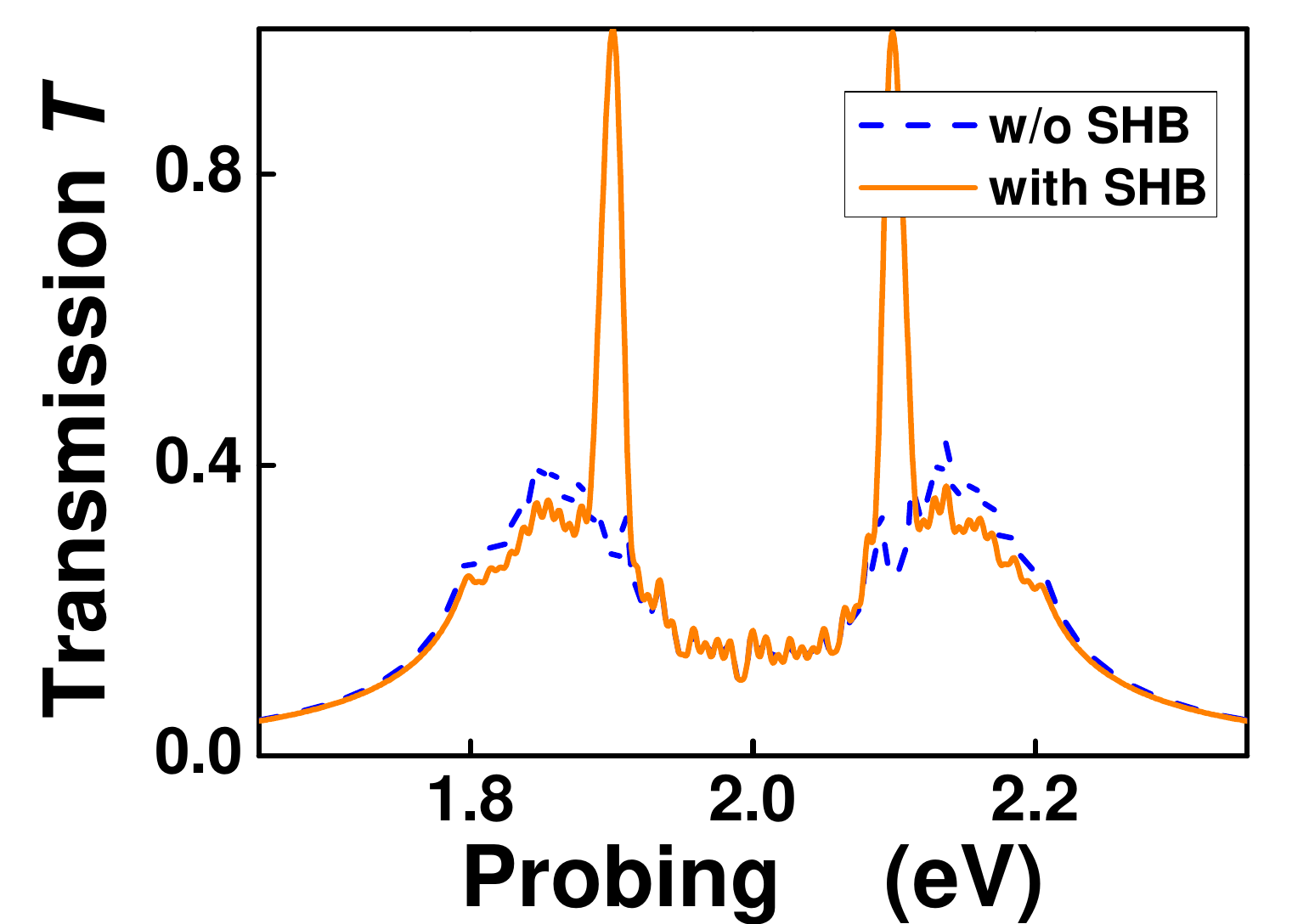}
\caption{The SHB effect under the condition of nonideal frequency comb.
The parameters used are kept the same
as those in Fig. 2(a) in the main text with $r =$ 0.5.
}
\label{figs1}
\end{figure}

\subsection{Randomly-distributed transition frequencies}

Next we consider another case where the transition frequencies of emitters are randomly distributed by the Lorentzian distribution and the couplings with plasmonic cavity are kept in constant for all the emitters.
Particularly, we sample $N=5000$ emitters from the same Lorentzian distribution and set the coupling strength of each emitter to be identical, $g_{i}=0.002$ eV.
It is found in Fig. \ref{figs2}(a) that the SHB effect can be observed in the dense emitter ensemble with individually weak coupling strength.

The SHB effect will be in stronger contrast to the background spectrum when the number of emitters becomes larger. As shown in Fig. \ref{figs2}(b), the SHB effect become more and more significant as the emitter number goes from 2000 to 6000. Here we show the case where the plasmonic cavity is coupled with an emitter ensemble with randomly-distributed transition frequencies.
It is found that the SHB effect will be in stronger contrast to the background spectrum when the number of emitters become larger.

\begin{figure}[h]
\includegraphics[width=9cm]{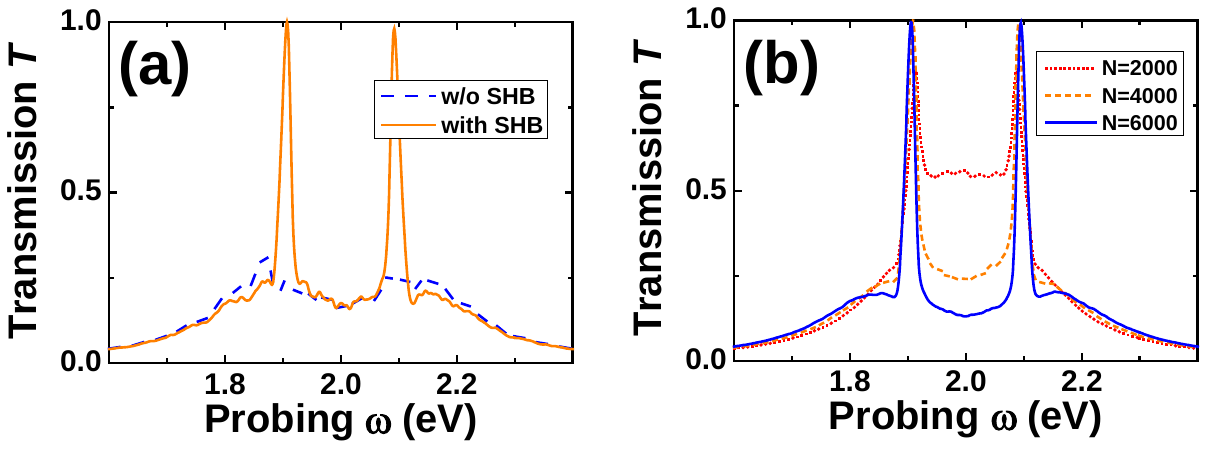}
\caption{(a) The SHB effect for a Lorentzian distributed ensemble with $N =$ 5000 emitters.
The spectral holes are burned at $\omega_{\textrm{L}} =$ 1.9 eV and $\omega_{\textrm{R}} =$ 2.1 eV with width of 0.033 eV.
(b) Transmission spectrum for SHB effect with different emitter numbers $N=2000,4000,6000$.
}
\label{figs2}
\end{figure}

\subsection{Different decay rates}
We also discuss the influence of fundamental decay rate to the SHB effect. For instance, when the decay rate of individual emitter $\Gamma_{i}$ increases from $0.01$eV to $0.05$ eV, the SHB effect will shrink gradually as seen in Fig. \ref{figs3}.

\begin{figure}[h]
\includegraphics[width=6cm]{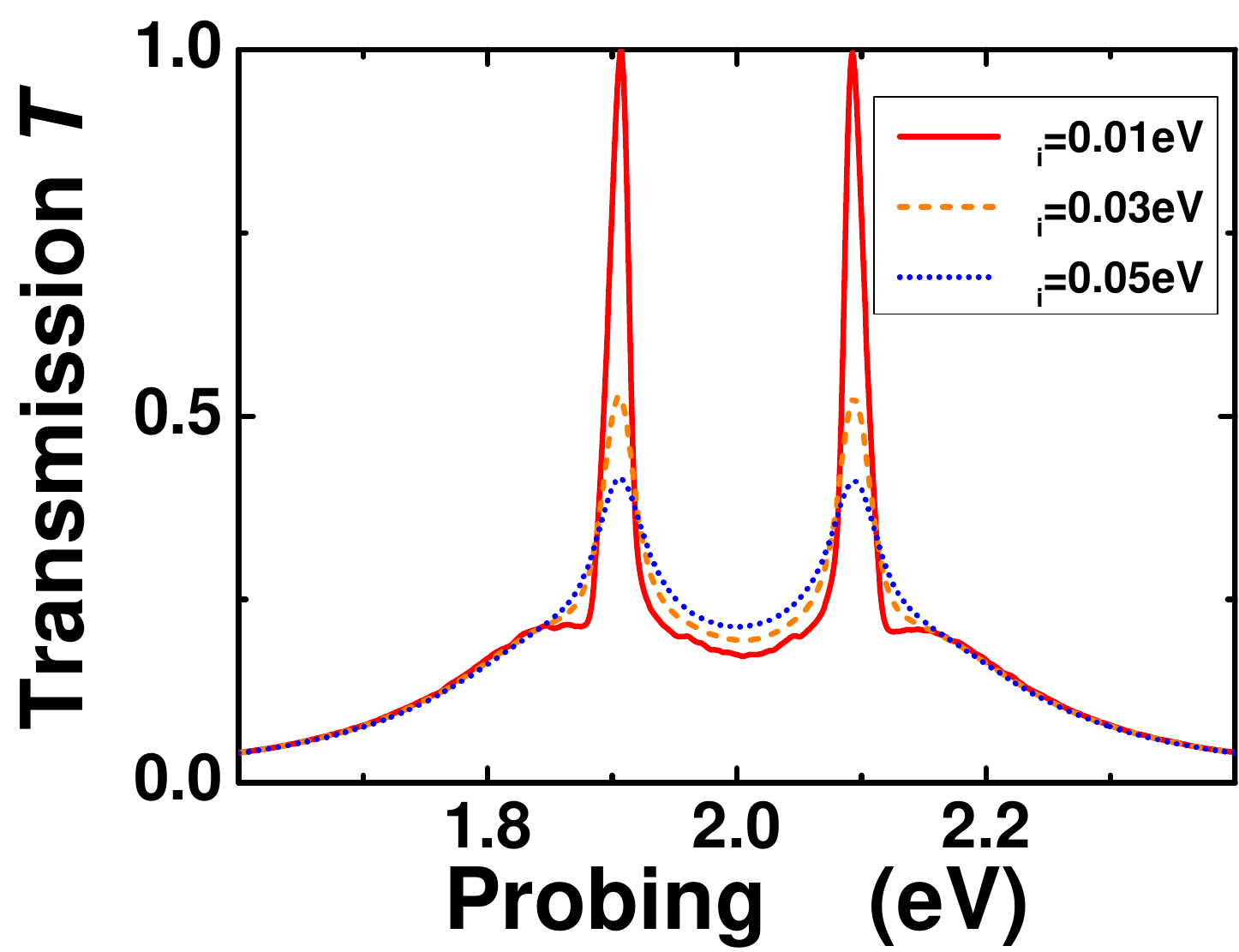}
\caption{The SHB effect with different decay rates $\Gamma_{i}=0.01,0.03,0.05$  eV.
The other parameters used are kept the same
as those in Fig. 2(a) in the main text.
}
\label{figs3}
\end{figure}

\section{Heating effects on plasmonic nanocavity}
\label{heating_time_domain}

\subsection{Heating effects of plasmonic nanocavity on SHB}

The burning pulse may induce local heating on the plasmonic metal nanoparticles, \textit{e.g.}, temperature effect \cite{yeshchenko2013temperature} or laser ablation effect \cite{doi:10.1002/adma.201103807,tarasenko2005laser,app9030363}, resulting in the changed properties of the plasmonic nanocavity ($i.e.$, $\omega_{\textrm{c}}$, and $\kappa$) during the hole burning process.
We study the heating effects on the plasmonic nanocavity and their impacts on SHB.
As indicated in Fig. \ref{figs4}(a), when the plasmon resonance changes to $\omega_{\textrm{c}}'$  (either red-shift or blue-shift with respect to original $\omega_{\textrm{c}}=2$ eV), the two SHB peaks become asymmetric.
The clear feature of Rabi oscillation will gradually disappear when such shift exceeds 120 meV as shown in Fig. \ref{figs4}(b), defining the critical limit to observe SHB if plasmonic nanocavity is changed.
On the other hand,  the impact from the changed decay rate $\kappa'$ seems less critical. As expected, increased $\kappa'$ results in two blunt SHB peaks and reduced Rabi oscillation as shown in Fig. \ref{figs5}.

\begin{figure}[h]
\includegraphics[width=12cm]{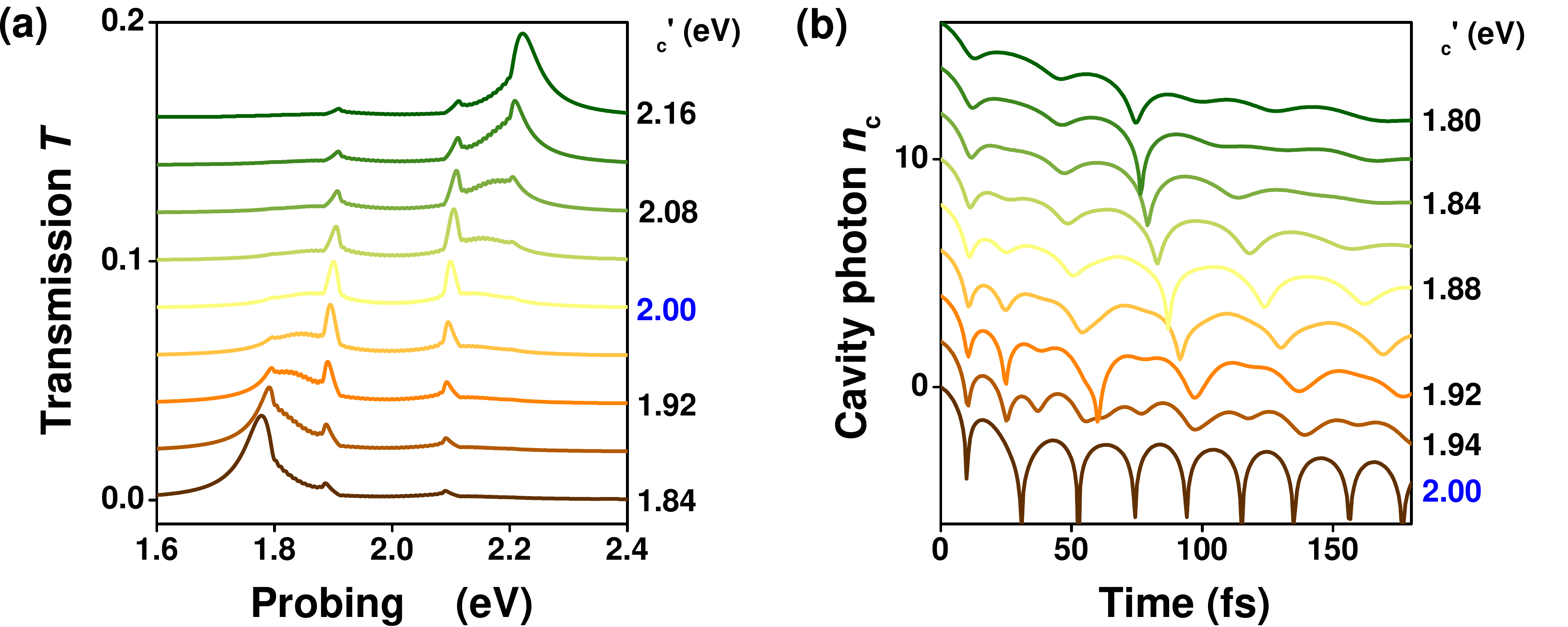}
\caption{The effect of the changed resonant frequency of plasmonic nanocavity $\omega_{\textrm{c}}'$ (from its original value of 2 eV before SHB) on the spectral hole burning effect.
The other parameters used are kept the same
as in Fig. 2(a) in the main text.
}
\label{figs4}
\end{figure}

\begin{figure}[h]
\includegraphics[width=12cm]{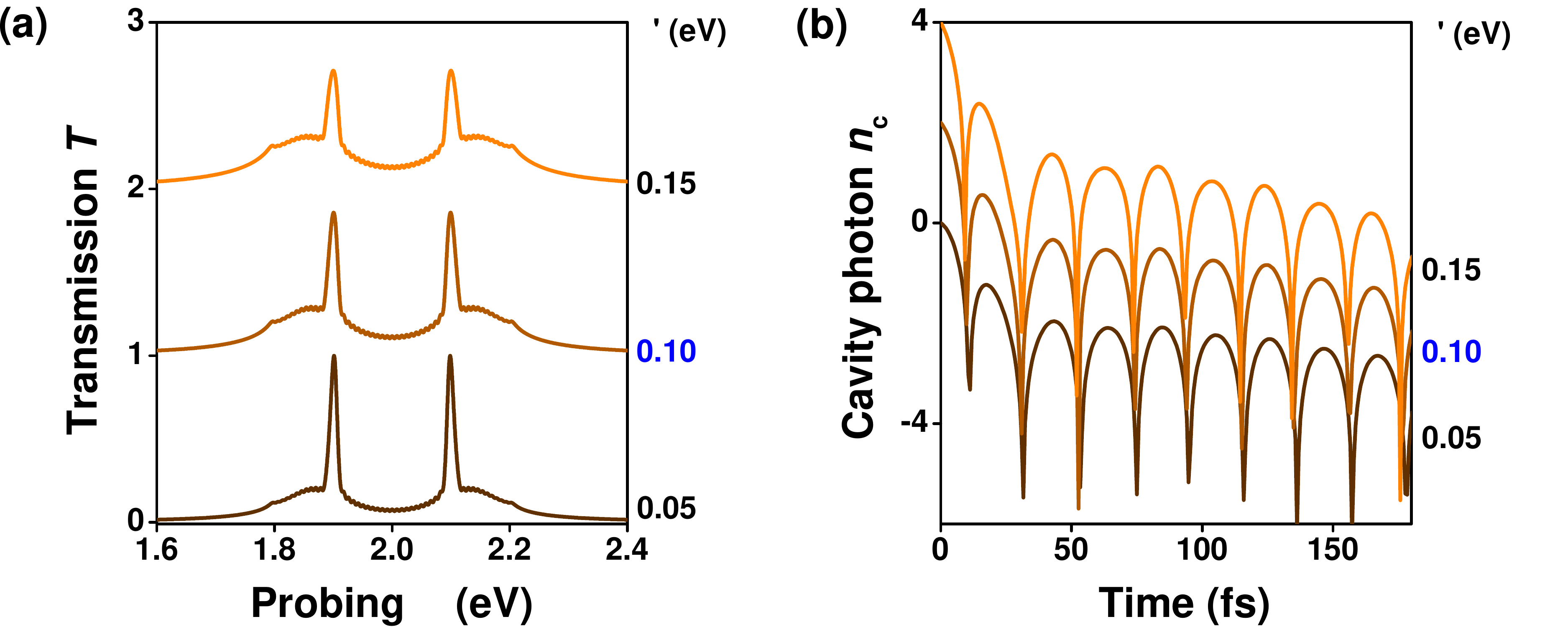}
\caption{The effect of the changed decay rate of plasmonic nanocavity $\kappa'$ (from its original value of 0.10 eV before SHB) on the spectral hole burning effect.
The other parameters used are kept the same
as those in Fig. 2(a) in the main text.
}
\label{figs5}
\end{figure}

\subsection{Full-wave optical simulations of plasmonic nanocavity}

In our full-wave optical modeling, we consider two closely-spaced Au nanospheres and solve the scattering problem for such subwavelength conductive nanostructures in an oscillating electromagnetic field \cite{wu2013fowler,huang2013synthesis}.
This is done by solving the full set of three-dimensional Maxwell's equations for the electric and magnetic fields using the finite element method.
The permittivity of Au is taken from the Johnson and Christy handbook \cite{johnson1972optical}.
In our simulations, we assume that: (i) the nanosphere has a diameter of 60 nm and placed closely to each other with a gap of $d=5$  nm;  (ii) the nanosphere dimer is embedded in an air environment (refractive index of 1); and
(iii) plane-wave excitation from the top with a background electric field $|E_{0}|=1$ V/m along the long axis of the nanosphere dimer.

Upon solving the electric and magnetic fields, the model calculates the spectrum of power absorption (\textit{i.e.}, the volume integration of the resistive heating) inside the Au nanospheres to identify the resonant wavelengths as shown in Fig. 3(b) in main text (symbols).
By plotting the spatial distributions of the calculated electric fields at the resonant wavelengths, we identify the plasmon resonance peak and fit it with a Lorentz curve (dotted lines) to extract the properties of the plasmonic nanocavity, resonant frequency $\omega_{\textrm{c}}$ and decay rate $\kappa$ ($i.e.$, the full width half maximum of the peak).
These parameters are then taken into the quantum simulation model to study the spectral hole burning effect.
All these full-wave optical calculations are performed based on the scattered-field formulation in the COMSOL multiphysics $-$ RF module, and a perfectly matched layer (PML) boundary is applied to eliminate the back reflections of the incident radiation.

\section{Dependence of Rabi oscillation amplitude on driving strength}
\label{dep_driving}

A driving laser field $E_{\textrm{L}}(t)$ with probing frequency $\omega$ pumps the entire system via the dipole moments of cavity $\mu_{\textrm{c}}$ and emitters $\mu_{\textrm{e}}$ with the strengths $\Omega_{\textrm{c}}(t)=\mu_{\textrm{c}}E_{\textrm{L}}(t)$ and $\Omega_{\textrm{e}}(t)=\mu_{\textrm{e}}E_{\textrm{L}}(t)$, where $\mu_{\textrm{c}}=19\mu_{\textrm{e}}$ and driving strength $\Omega_{\text{e}}=\Omega_{\text{c}}/19=1$ meV are used throughout the studies in main text.
For the scheme of $\pi$-phase-switched rectangular pulses, we can increase the electric field strength of the driving laser $E_{\text{L}}$ or the driving strength $\Omega_{\text{e}}=\Omega_{\text{c}}/19$ to amplify the Rabi oscillation amplitude. We find that the amplitude is proportional to the square of driving strength as shown in Fig. \ref{figs6}.

\begin{figure}[h]
\includegraphics[width=6cm]{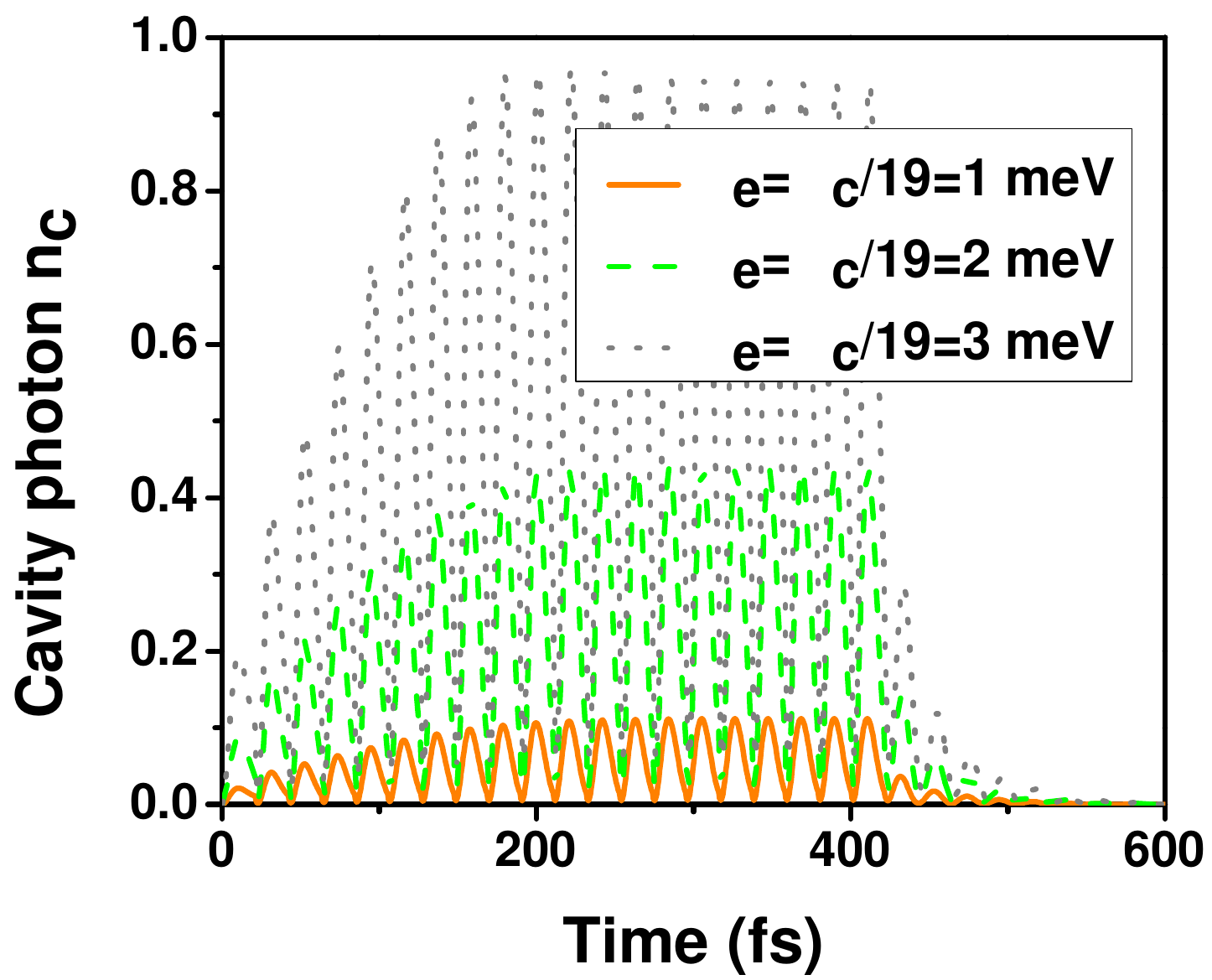}
\caption{Dependence of Rabi oscillation amplitude on the driving strength $\Omega_{\text{e}}=\Omega_{\text{c}}/19=$1, 2, and 3 meV. The other parameters used are kept the same as those in Fig. 4(b) in the main text.}
\label{figs6}
\end{figure}

\end{widetext}


%

\end{document}